\documentclass[fleqn,12pt]{article}
\usepackage{epsfig}
\newcommand{\be}{\begin{equation}}
\newcommand{\ee}{\end{equation}}

\newcommand{\eq}[1]{eq.~(\ref{#1})}

\newcommand{\pbar}{\bar p}
\def\signn{\sigma_{\rm nn}}
\def\siggp{\sigma_{\gamma\rm p}}
\def\siggg{\sigma_{\gamma\gamma}}
\def\rhonn{\rho_{\rm nn}}
\def\rhogp{\rho_{\gamma\rm p}}
\def\rhogg{\rho_{\gamma\gamma}}
\def\Bnn{B_{\rm nn}}
\def\Bgp{B_{\gamma\rm p}}
\def\Bgg{B_{\gamma\gamma}}
\begin{document}    
\renewcommand\thepage{\ }
%
%
\begin{titlepage} 
%
\newcommand\reportnumber{1000} 
\newcommand\mydate{April 4, 2002} 
\newlength{\nulogo} 
\settowidth{\nulogo}{\small\sf{N.U.H.E.P. Report No. \reportnumber}}
\title{\hfill\fbox{{\parbox{\nulogo}{\small\sf{Northwestern University: \\
N.U.H.E.P. Report No. \reportnumber\\ \mydate
}}}}\vspace{1in} \\
{Forward Compton Scattering, using Real Analytic Amplitudes}}
 
\author{Martin M. Block
\thanks{Work partially supported by Department of Energy contract
DA-AC02-76-Er02289 Task D.} \\
{\small\em Department of Physics and Astronomy,} \vspace{-5pt} \\ 
{\small\em Northwestern University, Evanston, IL 60208}\\
\vspace{.05in}\\
}
\vfill
\vspace{.5in}
\date{} 
\maketitle
\begin{abstract} 
\noindent We analyze forward Compton scattering, using real analytic amplitudes. By fitting the total $\gamma p$ scattering cross section data in the high energy region $5\ {\rm GeV}\le \sqrt s\le 200$ GeV, using a cross section rising as $\ln^2 s$, we calculate $\rho_{\gamma p}$, the ratio of the real to the imaginary portion of the forward Compton scattering amplitude, and compare this to $\rho_{nn}$, the ratio of the {\em even} portions of the $pp$ and $\pbar p$ forward scattering amplitudes.  We find that the two $\rho$-values are, within errors, the same in the 
c.m.s. energy region $5 \ {\rm GeV}\le\sqrt s\le 200$ GeV, as predicted by a factorization theorem of Block and Kadailov\cite{bk}.
\end{abstract}  
\end{titlepage} 
%
\pagenumbering{arabic}
\renewcommand{\thepage}{-- \arabic{page}\ --}  
%
\section{Introduction}
The $\rho$-value is defined as the ratio of the real to the imaginary portion of the forward scattering amplitude.   
Block and Kadailov\cite{bk} have shown that $\rho_{\rm nn}=\rho_{\gamma p}$ if one uses eikonals for the even portion of nucleon-nucleon scattering and $\gamma p$ scattering that have {\em equal} opacities, {\em i.e.,} eikonals that have the same value at impact parameter $b=0$. This is the equivalent of the more physical statement that
\be
\left(
\frac{\sigma_{\rm elastic}(s)}{\sigma_{\rm tot}(s)}
\right)_{\gamma p}
=\left(\frac{\sigma_{\rm elastic}(s)}{\sigma_{\rm tot}(s)}\right)_{nn},\quad {\rm for\  all\ }s.\label{sigratios}
\ee
Block and Kaidalov\cite{bk} have proved three  factorization theorems:
\begin{enumerate}
\item \label{enum:sig}  \[ \frac{\signn(s)}{\siggp(s)}=\frac{\siggp(s)}{\siggg(s)},\] where the $\sigma$'s are the total cross sections for nucleon-nucleon, $\gamma$p and $\gamma\gamma$ scattering,
\item \label{enum:B} \[ \frac{\Bnn(s)}{\Bgp(s)}=\frac{\Bgp(s)}{\Bgg(s)},\] where the $B$'s are the nuclear slope parameters for elastic scattering,  
\item \label{enum:rho}\[ \rhonn(s)=\rhogp(s)=\rhogg(s),\] where the $\rho$'s are the ratio of the real to imaginary portions of the forward scattering amplitudes,  
\end{enumerate}
with the first two  factorization theorems having their own proportionality constant.  These theorems are exact, for {\em all } $s$  (where $\sqrt s$ is the c.m.s. energy), and survive exponentiation of the eikonal\cite{bk}. The last theorem is valid independently of the model which takes one from $nn$ to $\gamma p$ to $\gamma\gamma$ reactions, as long as the respective eikonals have equal opacities, {\em i.e.,} \eq{sigratios} holds.   We wish to demonstrate experimentally here the validity of the theorem that states that $\rhonn(s)=\rhogp(s)$. However, no data are available in the hadronic sector for $\rhogp$.  The purpose of this note is to analyze forward Compton scattering at high energies in order to extract  $\rho_{\gamma p}$ and then to compare it to $\rho_{nn}$. 

Damashek and Gilman\cite{gilman} in 1970 have calculated $\rho_{\gamma p}$ using a singly-subtracted dispersion relation, up to a gamma ray laboratory energy $\nu=20$ GeV, {\em i.e.}, to a c.m.s. energy of $\sqrt s=6.2$ GeV. Here we extend the $\rho_{\gamma p}$ evaluation up to $\sqrt s=200$ GeV and then compare the results to $\rho_{nn}$, the ratio of the {\em even} portions of the $\pbar p$ and $pp$ forward scattering amplitudes (for references, see \cite{blockcr}).  In this paper we will calculate $\rho_{\gamma p}$ using real analytic amplitudes (see Section G. p.583 of ref. \cite{bc}).  It is shown in ref. \cite{bc} that the numerical complexities of dispersion relations in analyzing $\pbar p$ and $pp$ scattering can be circumvented by direct use of analytic functions to fit forward $\pbar p$ and $pp$ scattering amplitudes, a technique first proposed by Bourrely and Fischer\cite{bourrely}. We will introduce a variant of the Block and Cahn analysis\cite{bc} appropriate for Compton scattering, $\gamma p\rightarrow \gamma p$.
\section{Preliminaries}
This work largely follows the procedures and conventions used by Block and Cahn\cite{bc}. We use units where $\hbar =c=1$. The variable $s$ is the square of the c.m. system energy, whereas $\nu$ is the laboratory system momentum. In terms of the {\em even}  laboratory scattering amplitude $f_+$, where $f_+(\nu)=f_+(-\nu)$, the
total unpolarized  Compton cross section $\sigma_{\rm tot}$ is given by\cite{gilman}
\begin{eqnarray}
\sigma_{\rm tot}&=&\frac{4\pi}{\nu}{\rm Im}f_+(\theta=0),\label{optical}
\end{eqnarray}
where $\theta$ is the laboratory scattering angle.
We will assume that our amplitudes are real analytic functions with a simple cut structure\cite{bc}. 
For $\gamma p$ scattering, the assumed cut structure is a
left-hand cut that begins at the gamma ray energy $-\nu_0$ and a
right-hand cut that begins at the gamma ray energy $\nu_0$,
with a real amplitude  on the real axis between $-\nu_0$ and $\nu_0$. The threshold gamma ray energy for pion production
is $\nu_0=m_\pi+\frac{m_\pi^2}{m}\approx 0.16$ GeV, where  $m_\pi$ and $m$ are the pion and proton masses, respectively.  We will use an even amplitude for $\gamma p$ reactions in the high energy region $\nu>>\nu_0$, far above any cuts,  (see ref.\cite{bc}, p. 587, eq. (5.5a), with $a=0$), where the even amplitude simplifies considerably and is given by
\begin{equation}
f_+=i\frac{\nu}{4\pi}\left\{A+\beta[\ln (s/s_0) -i\pi/2]^2+cs^{\mu-1}e^{i\pi(1-\mu)/2}\right\}+C_{\rm subtraction},\label{evenamplitude_gp}
\end{equation}
where $A$, $\beta$, $c$, $s_0$ and $\mu$ are real constants. The additional real constant  $C_{\rm subtraction}=f_+(0)$ is the subtraction constant at $\nu=0$ needed in a singly-subtracted dispersion relation\cite{gilman} for the reaction $\gamma +p\rightarrow\gamma + p$ and is given by the Thompson scattering limit, {\em i.e.,} $f_+(0)=-\alpha/m=-3.03\  \mu {\rm b\  GeV}$.  In \eq{evenamplitude_gp}, we have assumed that the Compton cross section rises as $\ln^2 (s)$ at ultra-high energies. 

The real and imaginary parts of \eq{evenamplitude_gp} are given by
\begin{eqnarray}
{\rm Re}\frac{4\pi}{\nu}f_+ &=&\beta\,\pi \ln s/s_0-c\,\cos(\pi\mu/2)s^{\mu-1}+\frac{4\pi}{\nu} f_+(0)\label{real}\\ 
{\rm Im}\frac{4\pi}{\nu}f_+ &=&A+\beta\left[\ln^2 s/s_0-\frac{\pi^2}{4}\right]+c\,\sin(\pi\mu/2)s^{\mu-1}+, \label{imaginary}
\end{eqnarray}
where $s=2m\nu +m^2\approx 2m\nu$ and with $c$ being a real constant. Using equations ({\ref{optical}), (\ref{real}) and  (\ref{imaginary}), we find the total cross section for high energy Compton scattering is given by
\be
\sigma_{\rm tot}= A+\beta\left[\ln^2 s/s_0-\frac{\pi^2}{4}\right]+c\,\sin(\pi\mu/2)s^{\mu-1} , \label{sigmatot}
\ee 
and that $\rho$, the ratio of the real to the imaginary portion of the forward scattering amplitude, is given by
\be
\rho=\frac{\beta\,\pi\ln s/s_0-c\,\cos(\pi\mu/2)s^{\mu-1}+\frac{4\pi}{\nu} f_+(0)}{\sigma_{\rm tot}},\label{rho}
\ee 
with $f_+(0)=-3.03\  \mu {\rm b\  GeV}.$  
We will use units of $\nu$ in GeV and $s$ in GeV$^2$, and cross sections in $\mu$b. We have to fit the 5 real constants $A$, $c$, $\beta$, $s_0$ and $\mu$.  If we assume that the  term in $c$ is a Regge descending term, then $\mu=1/2$.  
 
The high energy behavior that was assumed by Damashek and Gilman\cite{gilman} in 1970 when they calculated $\rho_{\gamma p}$ numerically using dispersion relations was that the cross section approached a constant value asymptotically, {\em i.e.,} they assumed that $\sigma_{\rm tot}$, for $\nu \rightarrow \infty$, was given by
\be
\sigma_{\rm tot}=A+\frac{c'}{\nu^{1/2}}, \label{powerlaw}
\ee
with $A=96.6\mu {\rm b}$ and $c'=70.2\mu {\rm b\ GeV}^{1/2}$, with $\nu$ 
measured in GeV.  However, today we know  experimentally that the cross section rises at high energies and that the rising term can be fit by a $\ln^2 s$ term, as expressed in \eq{sigmatot}.
 
As a check on our real analytic amplitude analysis, the highest energy $\rho$ values of ref. \cite{gilman}, where Damashek and Gilman  used  dispersion relations assuming the asymptotic cross section of \eq{powerlaw}, can be simply reproduced 
from \eq{rho} with $\beta =0$
 and  \eq{powerlaw} by the relation
\be
\rho_{\rm analytic}=-\frac{70.2/\nu^{1/2}+38.07/\nu}{96.6+70.2/\nu^{1/2}}.\label{analyticrho}
\ee
The $\rho$-values calculated from \eq{analyticrho} and the results from the singly-subtracted dispersion relation of ref. \cite{gilman} agree to better than 2\% over the energy range $10\ {\rm GeV}\le \nu \le 20$ GeV.  The numerical agreement is excellent---the simplicity and ease of calculation at high energies using real analytical amplitudes compared to a dispersion relation analysis is clear.  
  
The $\rho$ values in ref. \cite{gilman} were only calculated up to $\nu=20$ GeV, which corresponds to a c.m.s. energy of $\sqrt s=6.2$ GeV. We  now make an amplitude analysis for $\rho_{\gamma p}$ using a {\em rising} cross section, asymptotically going as $\ln^2 s$ , by fitting the experimental cross sections $\sigma_{\rm tot}$ in the energy interval  
$5\ {\rm GeV}\le \sqrt s\le 200$ GeV to the parameters $A$, $c$, $\beta$ and $s_0$ of \eq{sigmatot}, using a Regge descending trajectory with $\mu=0.5$.
\section{Results and Conclusions}
Since cross sections for $\gamma p$ scattering are now available for c.m.s. energies up to 200 GeV, we made a $\chi^2$ fit to the experimental $\sigma_{\rm tot}(\gamma p)$ data in the c.m.s. energy interval $5 {\ \rm GeV}\le \sqrt s\le 200$ GeV.  We find a reasonable representation of the data using \eq{sigmatot}, with a $\chi^2$ per degree of freedom of 0.98 for 40 degrees of freedom, with the coefficients: 

$A= 115\pm 13$, $c=35.1\pm 97$, $\beta=1.07\pm 0.57$, $s_0=68.4\pm 80.2$ GeV$^2$, 

\noindent using a fixed value of $\mu=0.5$ (cross sections in $\mu$b when $s$ is in GeV$^2$). This fit, plotted as a function of c.m.s. energy, gives the dashed cross section curve $\sigma_{\rm tot}(\gamma p)$ in  Fig. \ref{fig:siggammap}, as well as the dashed  $\rho_{\gamma p}$ curve in Fig. \ref{fig:rhocompton}, using \eq{sigmatot} and \eq{rho}, respectively. Several remarks are in order. The experimental data are taken from the Particle Data Group compilation\cite{groom} and include the only three high energy points---in the neighborhood of 200 GeV---that are available.  The very large errors of $\beta$ and $s_0$ clearly reflect the fact that these few points also have large systematic errors ranging from $\approx 7$ to 21\%. The accuracy is sufficient to show that the cross section rises, but not much more.  Obviously, much more precise data at high energies are required for pinning down the $\log^2s$ parameters.   

In order to show visually the sensitivity of $\sigma_{\rm tot}(\gamma p)$ and $\rho_{\gamma p}$ to the parameters of the fit, we have also plotted in Fig. \ref{fig:siggammap} the dotted curve (a slight variation of the parameters of $A$ and $c$ within their errors), where we have set $A=107$ and $c=134$.  This curve has as its $\rho_{\gamma p}$ analog the dotted curve of Fig. \ref{fig:rhocompton}.  
\begin{figure}[p] 
\begin{center}
\mbox{\epsfig{file=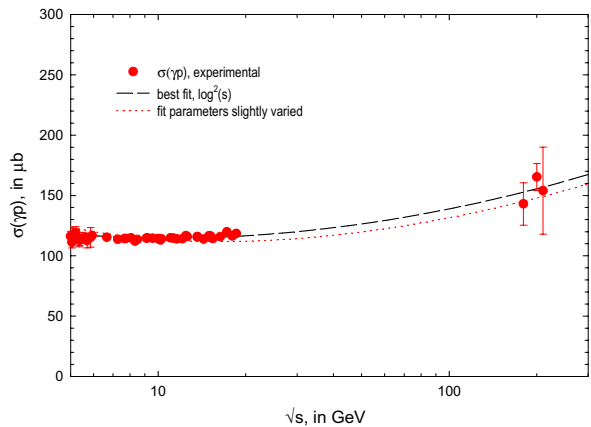,width=3.5in,%
bbllx=80pt,bblly=360pt,bburx=520pt,bbury=700pt,clip=}}
\end{center}
\caption[] {\footnotesize The dashed curve is $\sigma_{\rm tot}(\gamma p)$, the predicted total $\gamma p$ cross section from \eq{sigmatot}, using the central value parameters  $A= 115$, $c=35.1$, $\beta=1.07$, $s_0=68.4$ GeV$^2$, and $\mu=0.5$ of a $\chi^2$ fit, compared to the existing high energy experimental data in the c.m.s. energy interval $5 {\ \rm GeV}\le \sqrt s\le 200$ GeV.  The dotted curve varies the parameters slightly, with $A\rightarrow 107$ and $c\rightarrow 134$ (values within their errors). The corresponding $\rho_{\gamma p}$ curves are shown in Fig. \ref{fig:rhocompton}.}
\label{fig:siggammap}
\end{figure}
Using \eq{rho}, Fig. \ref{fig:rhocompton} shows our result for $\rho_{\gamma p}$  compared to $\rho_{nn}$, the $\rho$-value for nucleon-nucleon scattering found in ref. \cite{blockcr}, as a function of the c.m.s. energy $\sqrt s$, in GeV. The solid curve is $\rho_{nn}$; the dashed line  is the $\rho_{\gamma p}$ curve which corresponds to the central values $A= 115$, $c=35.1$, $\beta=1.07$, $s_0=68.4$ GeV$^2$; the dotted line is the $\rho_{\gamma p}$ curve which uses the slightly varied parameters $A=107$ and $c=134$.   
\begin{figure}[p] 
\begin{center}
\mbox{\epsfig{file=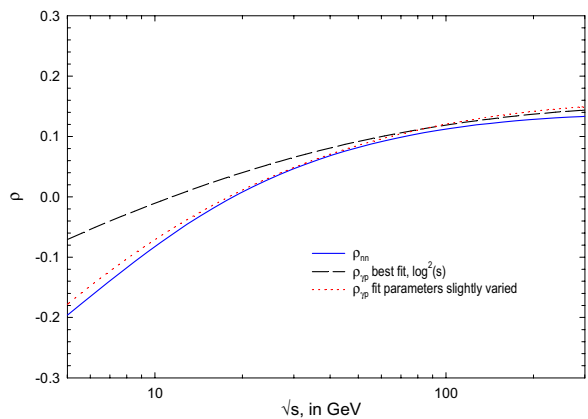,width=3.5in,%
bbllx=80pt,bblly=360pt,bburx=520pt,bbury=700pt,clip=}}
\end{center}
\caption[] {\footnotesize The solid curve is $\rho_{\rm nn}$, the predicted ratio of the 
real to imaginary part of the forward scattering amplitude for the
`elastic reactions , $\gamma +p\rightarrow V + p$ scattering
amplitude, where $V$ is $\rho$, $\omega$ or $\phi$ (using  factorization).
The dashed curve is $\rho_{\gamma p}$, the ratio of the real to imaginary
part of the forward scattering amplitude for Compton
scattering , $\gamma +p\rightarrow\gamma + p$, 
found from \eq{rho}, using real analytic amplitudes that asymptotically go as $\ln^2 s$, with best fit parameters $A= 115$, $c=35.1$, $\beta=1.07$, $s_0=68.4$ GeV$^2$, and $\mu=0.5$.  The dotted line is the $\rho_{\gamma p}$ curve  where the parameters of the fit have been slightly varied within their errors, with  $A\rightarrow 107$ and $c\rightarrow 134$.  The corresponding two curves for  $\sigma_{\rm tot}(\gamma p)$ are shown in Fig. \ref{fig:siggammap}.}
\label{fig:rhocompton}
\end{figure}
The agreement between the slightly modified $\rho_{\gamma p}$ and $\rho_{nn}$ over the energy interval $5\ {\rm GeV}\le \sqrt s\le 200$ GeV lends experimental support, in a model independent way, for the three factorization theorems of Block and Kadailov\cite{bk,bhp}. Of course, precision cross section data in the region $20\ {\rm GeV}< \sqrt s < 200$ GeV would enable us to strengthen this conclusion.
\section{Acknowledgments} The author would like to thank Professor J. D. Jackson for valuable critical comments and suggestions.

\end{document}